\documentclass[aps,preprint,showpacs]{revtex4}
\usepackage{graphicx}
\usepackage{dcolumn}

\begin{document}

\title{Dilepton production in proton-proton and quasifree proton-neutron 
reactions at 1.25 GeV}
\author{R. Shyam$^{1,2}$ and U. Mosel$^1$}
\affiliation {
$^1$ Institut f\"ur Theoretische Physik, Universit\"at Giessen, D-35392
Giessen, Germany \\ 
$^2$ Saha Institute of Nuclear Physics, Kolkata 700064, India }

\date{\today}

\begin{abstract}
We investigate the $pp \to ppe^+e^-$ and quasifree $pn \to pne^+e^-$ 
reactions within an effective Lagrangian model at the laboratory kinetic 
energy of 1.25 GeV for which experimental data have recently been reported 
by the HADES Collaboration. The model uses a meson-exchange approximation to 
describe the initial nucleon-nucleon ($NN$) scattering. Contributions to the 
reaction amplitudes are included from the NN bremsstrahlung as well as from 
the excitation, propagation and radiative decay of the $\Delta(1230)$ isobar 
state. It is found that the HADES data on the $e^+e^-$ invariant mass 
distribution in the $pp \to ppe^+e^-$ reaction are reproduced excellently by 
our model where the $\Delta$ isobar term dominates the spectrum. In the case 
of the quasifree $pn \to pne^+e^-$ reaction, a strong sensitivity to the pion 
electromagnetic form factor is observed which helps to bring the calculated 
cross sections closer to the data in the higher dilepton mass region.
\end{abstract}
\maketitle

\newpage
Dileptons ($e^+e^-$) provide a valuable tool to investigate the properties of 
the strongly interacting matter at high temperature and density formed in the 
relativistic and ultra-relativistic nuclear collisions because after their 
production they travel to the detectors almost undisturbed by the surrounding 
baryonic matter.  A recurring feature of the dilepton spectra measured in such 
collisions at low (DLS and HADES~\cite{por97,aga07}), intermediate [ super
proton synchrotron (SPS) \cite{aga05}] and high (PHENIX~\cite{afa07})
energies has been the significant 
enhancement observed in the intermediate dilepton mass region over the 
contributions from the electromagnetic decays of hadrons and long-lived mesons. 
While the major part of the excess yield seen at the SPS energies is attributed 
to the leptonic decay of the $\rho$ meson (formed in the $\pi^+\pi^-$ 
annihilation process) with strongly modified spectral function in the dense 
and hot hadronic matter~\cite{rap07}, at relativistic heavy ion collider 
(RHIC) energies it is believed to be 
more due the strong thermal contribution from the partonic phase (see, 
{\it e.g.}, Ref.~\cite{axe09}).

At lower beam energies (1-2 GeV/nucleon), various transport models
~\cite{bra98,ern98,rap00,she03} have been unable to explain the DLS data 
which must be attributed to some inherent problems in the theory as  the new 
measurements of the HADES Collaboration at these beam energies have confirmed 
the old DLS data~\cite{aga08}. Unlike the situation at high beam energies, 
the causes of this discrepancy - for  the light systems at least - are 
unlikely to be related to the in-medium effects. The insufficiently known 
cross sections for the dilepton production in elementary proton-proton ($pp$) 
and proton-neutron ($pn$) collisions, are an important reason behind this. 
Indeed, in a recent transport model calculation~\cite{bra08} it has been shown 
that if the input $pn$ bremsstrahlung cross sections (which are calculated 
within the soft-photon approximation model~\cite{gal87}), are scaled up in an 
adhoc manner by factors of 3-4, the observed dilepton yields of both DLS and 
HADES experiments at beam energies of 1-2 GeV/nucleon can be reproduced.

However, the microscopic models of dilepton production in elementary $NN$ 
reactions differ in their predictions of the $pn$ bremsstrahlung cross section. 
While the calculations performed within the effective Lagrangian models of 
Refs.~\cite{sch94,dej96,shy03,shy09} do not support the larger $pn$ 
bremsstrahlung yields, those of  the model of Refs.~\cite{kap06} favor the 
enhanced cross sections implemented in Ref.~\cite{bra08}. Therefore, 
to provide a reliable constraint for the dilepton yields in elementary 
reactions, the HADES Collaboration has very recently performed measurements 
for the dilepton production in not only the $pp$ reaction but also in the 
quasifree $pn$ reaction at 1.25 GeV beam kinetic energy~\cite{aga10}. The 
latter was measured by colliding a proton target with a deuteron beam of 
kinetic energy 1.25 GeV/nucleon and by detecting fast spectator protons from 
the deuteron breakup in a dedicated forward direction. In Ref.~\cite{aga10} 
the data on both $pp$ and quasifree $pn$ reactions were compared with the 
predictions of the model of Ref.~\cite{kap06} where it was noted that the 
calculations fail to describe the data for both reactions. While in the  
$pp$ case the dilepton yields were overestimated in the entire range of the 
dilepton invariant mass ($M$), those of the quasifree $pn$ reaction were 
overestimated (underestimated) at lower (higher) regions of $M$. 
 
The aim of this Rapid Communication is to investigate the  dilepton 
production in $pp$ and 
quasifree $pn$ reactions at the beam energy of 1.25 GeV within the effective 
Lagrangian model (ELM) of Refs.~\cite{sch94,shy03,shy09}. In order to compare 
our calculations with the HADES dilepton yields, we have also considered the 
following additional features: (i) for the quasifree $pn$ reaction the 
available energy in the center-of-mass (c.m.) system has been smeared to 
include 
the momentum distribution of the neutron in the deuteron using the Argonne V18
\cite{wir95} deuteron wave function. As a consequence, the $dp$ reaction 
results in a smeared $pn$ (quasifree) reaction where the available c.m. 
energies could be in excess of the threshold for the $\eta$ meson production 
(see, {\it e.g.}, Ref.~\cite{ing09}), (ii) because of this we have included 
the $\eta$ Dalitz decay cross sections in the total theoretical yields for 
the $pn$ reaction, and (iii) the contributions from the production and 
dileptonic decay of the subthreshold $\rho^0$ meson via the baryonic resonance
$N^*(1520)$, have been included for both $pp$ and quasifree $pn$ reactions.  
\begin{figure}
\centering
\includegraphics[width=.50\textwidth]{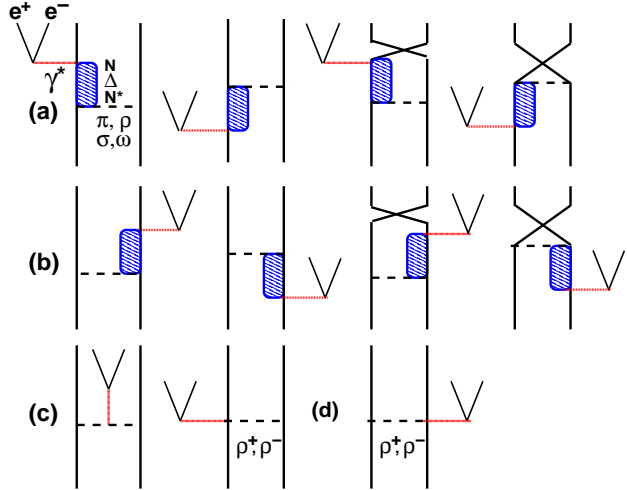}
\caption{
Graphical representation of our model to describe $e^+e^-$ production in 
$pp$ and $pn$ collisions. Diagrams (a) and (b) represent the $e^+e^-$
emission from an external nucleon line. (c) $e^+e^-$ emission from an internal
charged meson line in case of the $pn$ reaction. (d) diagrams representing
processes where $e^+e^-$ is emitted directly from the charged 
meson-nucleon-nucleon-photon vertices. 
}
\label{fig:Fig1}
\end{figure}

The Feynman diagrams [corresponding to both post emission and pre-emission 
(direct and exchange) processes] contributing to the dilepton production in 
the ELM are shown in Fig.~1. In case of the charged pion exchange 
(which happens for the $pn$ reaction) the internal meson line can also lead 
to dilepton emission (Fig.~1c). Initial interaction between two incoming 
nucleons is modeled by an effective Lagrangian which is based on the exchange 
of the $\pi$, $\rho$, $\omega$ and $\sigma$ mesons. The coupling 
constants at the nucleon-nucleon-meson vertices are determined by directly 
fitting the $T$ matrices of the $NN$ scattering in the relevant energy 
region~\cite{sch94}. These parameters are quite robust and have been used 
in successful descriptions of $NN \to NN\pi$ \cite{shy98}, $pp \to 
p\Lambda K^+$, $pp \to p\Sigma^0 K^+$~\cite{shy99,shy01} and $NN \to 
NN\eta$~\cite{shy07} reactions. 

The dilepton production proceeds via excitation, propagation and radiative 
decay of the intermediate nucleon or resonance states at either of the two 
colliding nucleon vertices (Figs.~1a and 1b). The nucleon intermediate 
states give rise to the $NN$ bremsstrahlung contribution. In calculations 
of various amplitudes we have used the same effective Lagrangians (and the 
corresponding parameters) for all of the hadronic and electromagnetic 
vertices as those given in Refs.~\cite{shy03,shy09}. We recall that for the 
$NN\pi$ vertex we have employed a pseudoscalar (PS) coupling where no 
derivative term of the pion field is involved. Therefore, no extra term 
(corresponding to a contact or the seagull diagram) appears in the model at 
the $NN\pi$ vertex when the electromagnetic coupling is included via the 
substitution $\partial^\mu \to \partial^\mu - iem A^\mu$, ($m$ is +1, 0, 
-1 for positive, neutral and negative pions). However, at the $NN\rho$ 
vertices such terms are always present as the corresponding coupling   
involve also the derivative of the meson field~\cite{ris84,hag89,hag91}. In 
any case, $\rho$-meson exchange terms contribute less than 5$\%$ to the total 
bremsstrahlung cross sections~\cite{shy10}. Thus, our calculations performed 
with a PS $NN\pi$ coupling, are almost free from the gauge invariance related
ambiguities which could be associated with the contact terms (see, {\it e.g.}, 
Ref.~\cite{uso05}). 

A note of caution should, however, be added here. With a PS $NN\pi$ coupling 
the role of negative energy states may be overestimated, whereas these are 
quite suppressed with the corresponding pseudovector (PV) 
interaction. Nevertheless, only the bremsstrahlung dilepton production 
amplitudes are expected to be  affected by the PS-PV coupling choice. 
Because, both $pp \to pp e^+e^-$ and $pn \to pn e^+e^-$ reactions are 
dominated by the delta isobar contributions at the beam energy of 1.25 GeV 
(as is shown in the following), our overall conclusions 
are not affected by the choice of the coupling at the $NN\pi$ 
vertex. In any case, the nucleon-antinucleon vertex may be considerably 
suppressed compared to the $NN$ vertex in presence of the timelike form 
factors~\cite{bro84}.

As the beam kinetic energy of 1.25 GeV is below the threshold of the $NN\eta$ 
channel (1.258 GeV), there is no contribution to the $pp \to pp e^+e^-$ cross 
sections from the $\eta$ Dalitz decay process. However, as explained earlier 
this can contribute to the quasifree $pn$ reactions. The $\pi^0$ Dalitz decay 
contributions must be taken into account as they dominate the cross sections 
at the lower ends of the dilepton invariant mass distributions in both the 
reactions.  

The $\eta$ Dalitz decay is treated as a two step process - the $\eta$ meson 
production by reactions $p + n \to p + n + \eta$  and $p + n \to d + \eta$, 
followed by the $\eta$-meson Dalitz decay. The total cross sections for
$\eta$ meson production reactions have been taken from Ref.~\cite{shy07} where 
a good description of the corresponding experimental data~\cite{cal98} is 
obtained. The $\eta$ Dalitz decay to $\gamma e^+e^-$ is calculated by using 
expressions given in Ref.~\cite{lan85}. A similar procedure is applied for the 
$\pi^0$ case where the production cross sections have been taken from 
Ref.~\cite{shy98} while for its Dalitz decay the formulas of Ref.~\cite{lan85} 
have been utilized. 

In calculations of dilepton yields from the production and decay of the 
subthreshold $\rho^0$ meson via the baryonic resonances, we consider only 
the $N^*(1520)$ resonance - other higher lying resonances are expected to 
contribute negligibly at the beam energy considered in this Rapid 
communication~\cite{bra99}.
We suppose this reaction to proceed as a $NN \to RN \to \rho^0NN \to e^+e^-NN$
process ($R$ represents a resonance), which leads to the following 
factorization of the cross section,   
\begin{eqnarray}
\frac{d\sigma(s,M)}{dM}^{NN \to NN e^+e^-} & = &
             \frac{d\sigma(s,M)}{dM}^{NN \to \rho^0 NN}\nonumber 
   \frac{\Gamma_{\rho^0 \to e^+e^-}(M)}{\Gamma_\rho^{tot}(M)},
\end{eqnarray}   
where $s$ is the square of the invariant mass associated with the incident 
channel. In Eq.~(1) the first term on the right hand side represents the 
differential cross section for the $\rho^0$ meson production in $NN$ collisions
which is calculated by following the procedure described in 
Refs.~\cite{fri97,pet98} using the same parameters as those given in 
Ref~\cite{pet98}. The second term is the branching ratio for the 
$\rho^0 \to e^+e^-$ decay.
$\Gamma_{\rho^0 \to e^+e^-}$ is the decay width of the $\rho^0$ meson to the 
dilepton channel, which is calculated within a strict vector meson dominance 
model as in Ref.~\cite{bra99}. $\Gamma_\rho^{tot}$ is the total $\rho$-meson 
width which is given by~\cite{man92} 
\begin{eqnarray}
\Gamma_\rho^{tot}(M) & = & \Gamma_{\rho^0 \to \pi \pi} \frac{r_C^2 k^3}
                         {M(1+r_C^2k^2)},
\end{eqnarray}   
where $k^2 = M^2/4 - m_\pi^2$. The parameter $r_C$ represents an 
interaction radius which is taken to be 2 fm. $\Gamma_{\rho^0 \to \pi \pi}$ 
= 0.150 $GeV$.  Eq.~(2) represents the partial width for the $\rho$ meson 
decay to the $2\pi$ channel only which has a branching ratio of nearly 
100$\%$. In our calculations, however, we have added to it also the width  
$\Gamma_{\rho^0 \to e^+e^-}$.  

In Fig.~2(a), we compare the calculated and the measured dilepton invariant
mass ($M$) distributions for the $pp \to ppe^+e^-$ reaction at the beam 
kinetic energy of 1.25 GeV. We recall that for this reaction only diagrams 
1(a) and 1(b) contribute. The theoretical cross sections have been folded 
with the appropriate detector acceptances provided to us by the HADES 
Collaboration~\cite{ing10}. We see that the total cross section (solid line)
(obtained by the coherent summation of $NN$ bremsstrahlung and $\Delta$ 
isobar amplitudes which will be referred as QM) for this reaction is 
dominated by the $\Delta$ isobar terms (dashed line). The $NN$ 
bremsstrahlung contributions (dashed-dotted line) are smaller by almost an 
order of magnitude for lower values of $M$ and by factors of 3-5 at higher 
$M$. The region of $M < 0.15$ GeV/$c^2$ is dominated by the $\pi^0$ Dalitz 
decay cross sections. 

It clear that our QM cross sections for the $pp \to ppe^+e^-$ reaction are 
in excellent agreement with the HADES data for $M$ $>$ 0.15 GeV/$c^2$. In 
contrast to this, the model of Ref.~\cite{kap06} overestimates the data 
everywhere in this region (see Fig.~1 of Ref.~\cite{aga10}). The similar 
observation was also made in comparisons with the DLS $pp$ dilepton data at 
1.04 GeV beam energy in Ref.~\cite{shy09}.
\begin{figure}
\centering
\includegraphics[width=.45\textwidth]{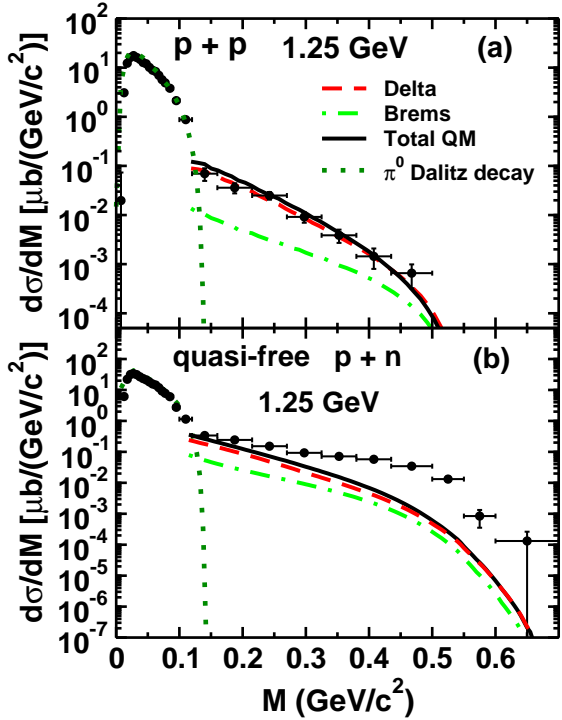}
\caption{[color online]
(a) The invariant mass distribution of the dileptons produced in the
$pp \to ppe^+e^-$ reaction at the beam energy of 1.25 GeV. The maximum
allowed value of $M$ is 0.545 GeV for this beam energy. (b) The same 
for the quasifree $pn \to pne^+e^-$ reaction at the same incident energy. 
The Fermi smearing as discussed in the text, has been applied to all the 
theoretical calculations shown in Fig. 2(b). Experimental data are from the
Ref.~\protect\cite{aga10}.
}
\label{fig:Fig2}
\end{figure}
Since this reaction is dominated by the $\Delta$ contributions, the larger 
$\Delta$ cross sections of Ref.~\cite{kap06} as compared to those of our model 
are the most likely reason for the differences seen in the predications of the 
two models.  

In Fig.~2(b), we show the same for the quasifree $pn \to pne^+e^-$ reaction 
at 1.25 GeV beam kinetic energy. Because of the Fermi smearing, the tails of 
various contributions extend to $M$ values larger than those of the $pp$ 
reaction which is in agreement with the data. However, the shape of the 
$pn$ spectra differs significantly in several ways from that of the $pp$ case 
in the region of $M$ beyond that dominated by the $\pi^0$ Dalitz decay process. 
Firstly, the $NN$ bremsstrahlung contribution now is relatively larger 
although the $\Delta$ isobar term still dominates the total cross section. 
Secondly, the QM cross sections significantly underpredict the HADES quasifree 
$pn$ data for $M$ $>$ 0.20 GeV/$c^2$ which is in sharp contrast to the $pp$ 
case. The difference between theory and the data varies from factors of 2-3 at 
the lower mass values to more than an order of magnitude for $M$ around 
0.5 GeV/$c^2$. It is important to understand this discrepancy between 
calculations and the data for the quasifree $pn \to pne^+e^-$ reaction as 
the dilepton excess in the intermediate mass range of 0.15 $< M (GeV/c^2)< $ 
0.60 observed in the $C + C$ collisions at 1 and 2 GeV/nucleon, can be 
explained by a superposition of experimental elementary $pp$ and $pn$ 
reactions as in-medium effects are almost negligible~\cite{aga10}. 

In none of the results shown so far (in Figs.~2a and 2b as well as in Refs.
\cite{shy03,shy09}) electromagnetic form factors were 
considered at any of the vertices. However, in the earlier work reported in 
Ref.~\cite{sch94} it was shown that the hadronic electromagnetic form factors 
have a significant influence on the dilepton spectra. Therefore, we now 
include the pion electromagnetic form factor (PEFF) [$F_\pi (M^2)$] at the 
charged internal meson line (Fig.~1c). In the present exploratory study we 
use the same form factors at the pion and nucleon vertices in order to preserve 
gauge invariance~\cite{sch94,hag91}. We have used two parameterizations for 
$F_\pi (M^2)$. The first one (to be referred as FF1) is written as 
\begin{eqnarray}
F_\pi (M^2) & = & \frac{m_\rho^2}{m_\rho^2 - M^2 - im_\rho \Gamma_\rho(M^2)},
             \nonumber
\end{eqnarray}
where $m_\rho$ is the $\rho$ meson mass and $\Gamma_\rho$ is the width for 
$\rho \to \pi \pi$ decay. The assumption inherent in FF1 is that the photon 
couples to the pion only via the $\rho^0$ meson. It reproduces the 
main features of the pion EFF both in time- and space-like regions
(see, Ref.~\cite{eri88}). The other, to be referred as FF2, is described 
extensively in Ref.~\cite{bro86},
\begin{eqnarray}
F_\pi (M^2) & = & \frac{0.4}{1-M^2/\lambda^2} + \frac{0.6}{1-M^2/2m_\rho^2}
      \frac{m_\rho^2}{m_\rho^2 - M^2 - im_\rho \Gamma_\rho(M^2)},
      \nonumber
\end{eqnarray}
where $\lambda^2$ = 1.9 $GeV^2$. The width $\Gamma_\rho(M^2)$ appearing in  
both FF1 and FF2 has been calculated by following the expressions given in 
Ref.~\cite{bro86}. FF2 is derived from the assumption that photon couples  
about 50$\%$ directly to the intrinsic quark structure of the pion and 
remaining 50$\%$ indirectly through the $\rho^0$ meson. FF2 provides a better 
description of the PEFF in the timelike region. The imaginary parts of both
FF1 and FF2 are proportional to two pion phase space - blow two pion 
production threshold both FF1 and FF2 are real. It should be stressed here 
that we have put both form factors on the mass shell. Given the high 
virtuality of the internal pions, the form factors should be functions of 
both pion momentum and the momentum transfer. However, the knowledge about 
the off-shell pion form factor is still scanty - the recent extraction the 
PEFF from the JLab electroproduction experiments provide information about 
essentially the on shell pion form factors only (see, e.g., Ref.~\cite{tad07}).
Therefore, we use the on shell PEFF in these calculations with a caveat that 
the off-shell PEFF could be larger than the on-shell one~\cite{brs01}. 
  
\begin{figure}
\centering
\includegraphics[width=.45\textwidth]{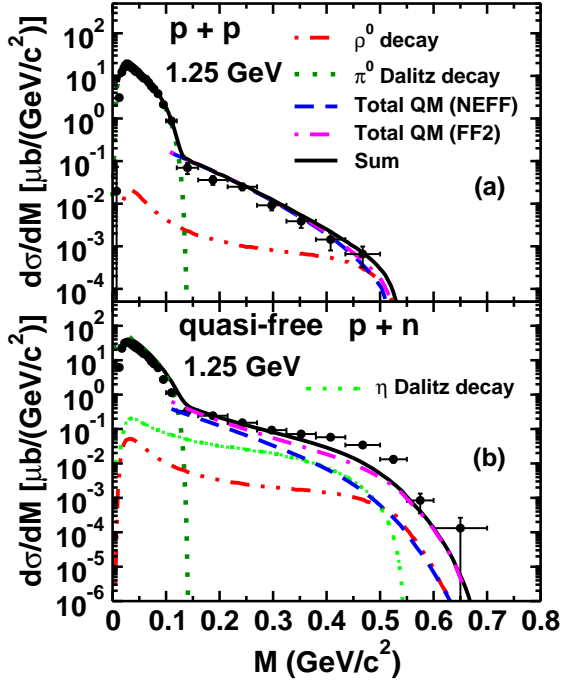}
\caption{[color online] 
(a) and (b) show the same reactions as in Figs.~2(a) and 2(b) but with 
electromagnetic form factors included at pion and nucleon vertices. Also shown 
are the contributions of meson Dalitz decays and subthreshold $\rho^0$ decay 
processes. Total QM cross sections obtained with (FF2) and without (NEFF) 
electromagnetic form factors are shown by dashed and dashed-dotted lines, 
respectively. The simple sum of the meson Dalitz decays, $\rho^0$ decay and 
full quantum mechanical (with FF2 form factors) cross sections are shown by 
the full line. The data are taken from the Ref.~\protect\cite{aga10}.
}
\label{fig:Fig3}
\end{figure}

In Fig~3(a), we show the effect of the PEFF and the contributions of the 
$\pi^0$ Dalitz decay and subthreshold $\rho^0$ decay process for the 
$pp \to ppe^+e^-$ reaction. We note that for this case, the introduction 
of the electromagnetic form factors (which are assumed to be the same for 
proton and pion vertices) makes hardly any difference to  results obtained 
without them. The effect of FF2 type of PEFF is barely observed only at 
the extreme end of the spectrum. Results obtained with the FF1 form factor 
are not shown here - there are even more closer to no PEFF results. 
Furthermore, the subthreshold $\rho$ decay cross sections too are of some 
relevance only in the extreme tail region. 
 
In Fig.~3(b), we show the total QM cross sections obtained without (NEFF) 
and with (FF2) electromagnetic form factor of FF2 type for the quasifree 
$pn \to pne^+e^-$ reaction (where Fig.~1c also contributes) at the beam 
energy of 1.25 GeV. We have not shown explicitly the cross sections 
obtained with FF1 type of PEFF in order not to overcrowd the figure - they 
lie between the NEFF and FF2 results. The larger cross sections obtained 
with FF2 form factors as compared to those with FF1 can be traced back to 
the fact that in the timelike region the former is significantly larger 
than the latter~\cite{eri88,bro86}. We note that with FF2 type of PEFF, 
the QM cross sections are significantly enhanced for $M >$ 0.3 GeV/c$^2$ 
and are larger than $\eta$ and $\rho^0$ decay contribution by almost an 
order of magnitude at larger values of $M$. The $\eta$ Dalitz decay 
contributions drop off strongly for $M$ beyond 0.50 GeV/$c^2$ due to phase 
space restrictions. In this region the $\rho^0$ decay cross sections become 
relatively stronger. It is seen that the simple sum of the QM (with FF2) and 
the meson decay cross sections is able to reproduce the data now for $M$ up 
to $\simeq$ 0.4 GeV/$c^2$ and for $M>$ 0.55 GeV/$c^2$. It should, however be 
stressed that there is a danger of double counting by explicitly including 
$\rho$ meson production and decay terms together with the form factor FF2 
which implicitly includes a $\rho$ meson bump. However, because the 
contributions of the explicit $\rho^0$ meson production process are 
relatively quite small as compared to that of the form factor FF2, this 
problem may not be too serious.

We remark that the final state interaction (FSI) effects ($pn$ and $pp$)
estimated within the Watson-Migdal method increase the magnitudes of the 
cross sections with increasing $M$ value. However, even at the extreme 
kinematical limits the FSI related enhancements in the cross sections are 
not more that 15-20$\%$ for these reactions at 1.25 GeV. This result is in 
agreement with those of Ref.~\cite{kap06}. Furthermore, the deuteronlike 
final states have also not been considered because the HADES measurements 
have ruled out such states in their data.
 
In summary, we extended our effective Lagrangian model for dilepton 
production in nucleon-nucleon collisions by including the pion 
electromagnetic form factors at the internal meson line in a way that still 
preserves gauge invariance and employed it to describe the 
new data of the HADES Collaboration for these reactions.

For the quasifree $pn \to pne^+e^-$ reaction, the inclusion of the 
electromagnetic form factors enhances significantly the cross sections 
for dilepton masses larger than 0.3 GeV/c$^2$. This is for the first time 
that dilepton production data in elementary proton-neuron reactions is 
shown to be so sensitive to the pion electromagnetic form factors. Although 
this effect was already noted in the early work of Ref.~\cite{sch94} but it 
could not be affirmed at that time because of the absence of data on the 
elementary $pn$ process. We find that the simple sum of the $\pi^0$ Dalitz 
decay and the ELM cross sections is able to describe the experimental 
invariant mass distribution of the dileptons everywhere except for the 
three points lying between 0.40 -0.55 GeV/$c^2$. The $\eta$ Dalitz and 
$\rho^0$ decay processes are of only minor consequence.  

For the $pp \to ppe^+e^-$ reaction, the ELM which remain almost unaffected by 
the electromagnetic form factors, provides on its own a good description of 
the data for dilepton invariant mass $>$ 0.15 GeV/$c^2$. This is in a marked 
contrast to the results shown in Ref.~\cite{aga10} where the model of 
Ref.~\cite{kap06} is found to grossly overestimate the data in this region. 
The simple sum of ELM cross sections and those of the meson decay processes 
provides an excellent description of the data in the entire region of the 
dilepton invariant mass. 

The authors are thankful to Ingo Fr\"ohlich and Tetyana Galatyuk for their 
help in implementing the acceptances of the HADES detector in our theoretical 
calculations.  This work is supported by the Federal Ministry of Education 
and Research (BMBF). One of the authors (RS) acknowledges support from the 
Helmholtz International Center for FAIR under the L\"OWE program.

\end{document}